\pdfoutput=1
\documentclass[pra,twocolumn,showpacs,floatfix]{revtex4-1}
\usepackage{times,amsmath,amssymb,amstext,latexsym,float,graphicx,color,ulem}
\usepackage{hyperref}

\hypersetup{colorlinks=true, citecolor=blue, urlcolor=blue, linkcolor=blue}

\begin{document}

\title{Droplet-superfluid compounds in binary bosonic mixtures}

\author{M.~Nilsson Tengstrand}
\email{mikael.nilsson\_tengstrand@matfys.lth.se}
\author{S.M.~Reimann}
\affiliation{Division of Mathematical Physics and NanoLund, Lund University, SE-221 00 Lund, Sweden }

\date{\today}

\begin{abstract}
While quantum fluctuations in binary mixtures of bosonic atoms with short-range interactions can lead to the formation of a self-bound droplet,
for equal intra-component interactions but an unequal number of atoms in the two components, there is an excess part 
that cannot bind to the droplet. Imposing confinement, as here through periodic boundary conditions in a one-dimensional setting,
the droplet becomes amalgamated with a residual condensate. 
The rotational properties of this compound system reveal simultaneous rigid-body and superfluid behavior in the ground state and uncover that 
the residual condensate can carry angular momentum even in the absence of vorticity. 
In contradiction to the intuitive idea that the superfluid fraction of the system would be entirely made up of the excess atoms not bound by the droplet, we find evidence that this fraction is higher than what one would  expect in such a picture. Our findings are corroborated by an analysis of the elementary excitations in the system, and shed new light on the coexistence of localization and superfluidity.
\end{abstract}

\maketitle

In the field of ultra-cold atomic quantum gases it was suggested early-on~\cite{Bulgac2002,Bedaque2003,Hammer2004} that 
quantum effects beyond mean-field (BMF) may lead to self-bound droplets of fermionic or bosonic 
atoms. For weakly-interacting single-component Bose gases, quantum fluctuations alone are often too small  
to play any significant role. Nevertheless, for binary or dipolar Bose gases the interactions 
may be adjusted such that the different main contributions to the mean-field (MF) energy nearly cancel out, leaving only a small residual term that can be tuned to equilibrate with the BMF part of the total energy.  A self-bound dilute boson droplet may then form~\cite{Petrov2015,Petrov2016}, with curious properties originating from its genuine quantum many-body nature.  Although originally proposed  for binary Bose gases~\cite{Petrov2016}, the first experimental observations of droplets stabilized by the Lee-Huang-Yang (LHY) quantum fluctuations~\cite{Lee1957} were made for strongly dipolar atoms such as Dysprosium~\cite{Kadau2016,Ferrier-Barbut2016,Ferrier-Barbut2016b,Schmitt2016} and Erbium~\cite{Chomaz2016}. Here, a  scenario similar to the binary case~\cite{Staudinger2018} 
arises due to the peculiarities of the dipolar interactions~\cite{Wachtler2016a,Wachtler2016b,Macia2016,Bisset2016,Baillie2017}.
Experiments with binary bosonic mixtures of potassium~\cite{Cabrera2018,Semeghini2018,Cheiney2018} or hetero-nuclear 
mixtures~\cite{Errico2019,ZGuo2021} followed soon after (for brief reviews on droplet formation, see Refs.~\cite{Bottcher2021,Luo2021}).
In low-dimensional systems, quantum fluctuations may be enhanced, facilitating and stabilizing the droplet formation process~\cite{Petrov2016,Astrakharchik2018,Parisi2019}, and 
the dimensional crossover has been discussed in Refs.~\cite{Zin2018,Ilg2018,Lavoine2021}. 
Recent work on binary self-bound states in 1D or quasi-1D also investigated corrections beyond LHY~\cite{Ota2020a},  applied the quantum Monte-Carlo method~\cite{Parisi2019,Parisi2020}, used the Bose Hubbard model~\cite{Morera2020,Morera2021a} or formulated an effective quantum field theory~\cite{Chiquillo2018}.  
It has been shown that microscopic pairing or dimer models agree with variational approaches~\cite{Hu2020c,Morera2021b}.  
Collective excitations~\cite{Cappellaro2018,Tylutki2020} and thermodynamic properties~\cite{Ota2020a,DeRosi2021} were also studied.\\
Droplets may also form in systems with inter-component asymmetry. One obvious realization is heteronuclear mixtures, see {\it e.g.} Refs.~\cite{Ancilotto2018,Errico2019,Minardi2019,ZGuo2021,Mistakidis2021}.  Another interesting scenario arises when the intra-component interactions are equal, but the components have different numbers of particles.  Then, the excess particles in the larger component cannot bind to the droplet~\cite{Petrov2015}, but instead form a uniform background into which the droplet is immersed~\cite{Mithun2020}.
In a similar line of thought, but for non-equal interactions within the components, it was recently suggested~\cite{Naidon2021}  
that a  mixed phase may coexist with a non-amalgamated gaseous component, where partial miscibility is caused by BMF contributions leading to  so-called  “bubble” phases with similarities to the droplet self-bound states. 

\begin{figure}[H]
\centering
\includegraphics[width = \columnwidth]{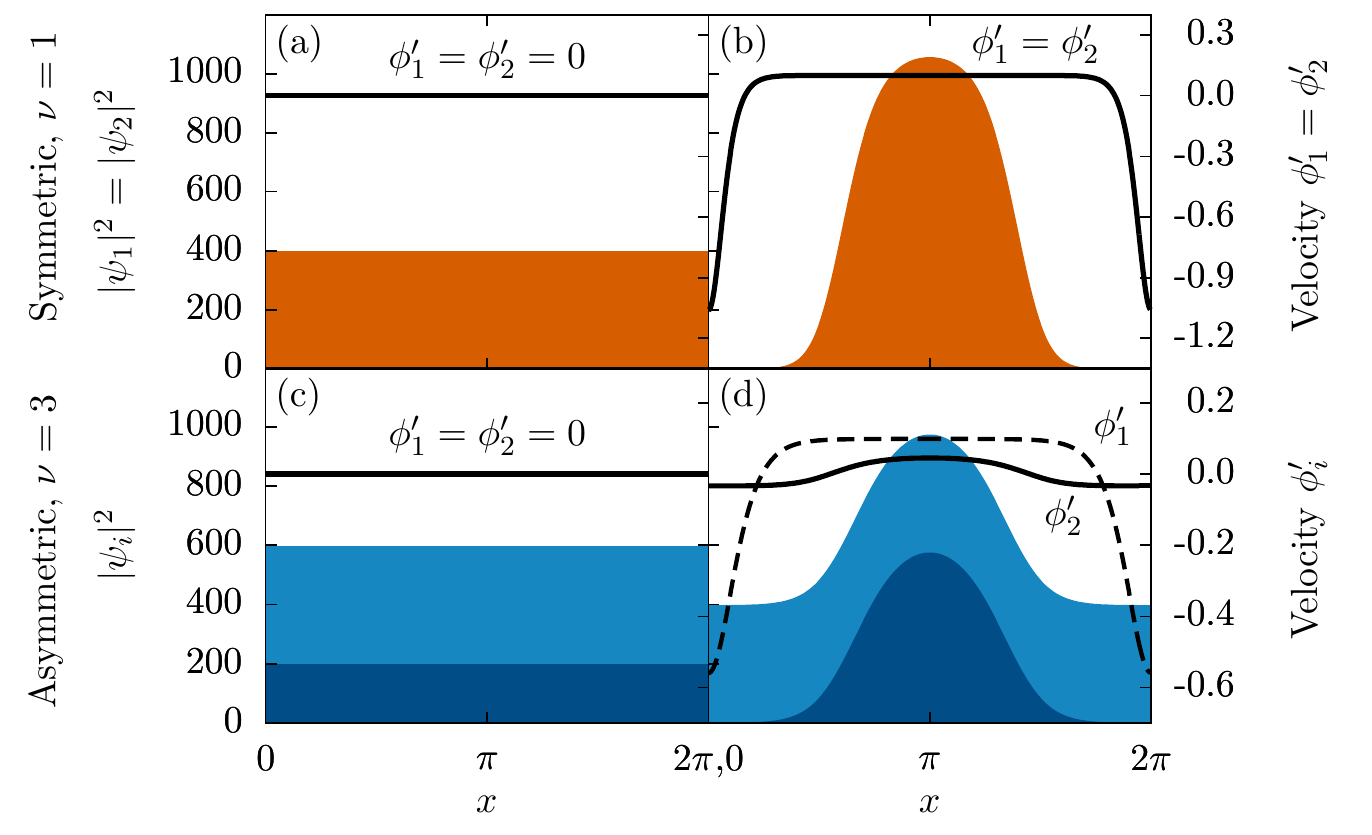}
\caption{({\it Color online}) Condensate densities $\vert \Psi _i\vert ^2$ {\it (red, blue)} and velocities $\phi _i^{\prime }$ {\it (black lines)} of the two components in the rotating frame where $\Omega =0.1$.  
The {\it upper panel} shows the symmetric case, $\nu =1$. (a) $\vert \Psi _1\vert ^2=\vert \Psi _2\vert ^2$ for $g=0.5$  {\it (red)}, being homogeneous along the ring. (b) as in (a) but for $g=1.2$, showing droplet formation. The {\it lower panel} shows the asymmetric case for $\nu =3$ for the same parameters as in the upper panel.
The axis to the right in both panels shows the corresponding condensate velocities $\phi _i^{\prime }$.}
\label{fig:dens}
\end{figure}

\begin{figure}[H]
\centering
\includegraphics[width =0.9\columnwidth]{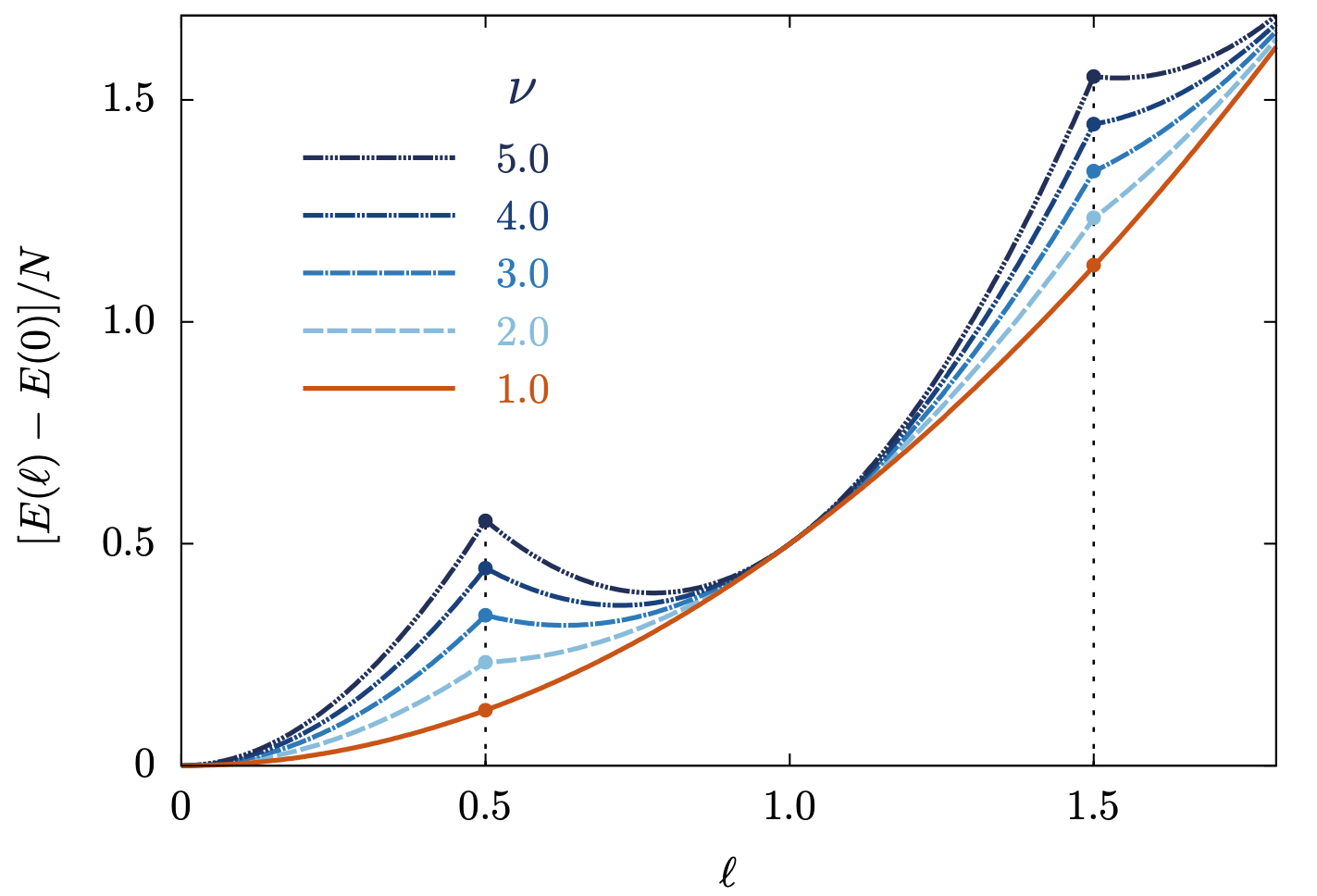}
\caption{({\it Color online}) Energy as a function of  angular momentum per particle for $\nu = 1,2,3,4,5$ as indicated by the legend, where $g = 1.5$.}
\label{fig:disp}
\end{figure}

\begin{figure}[H]
\centering
\includegraphics[width = 0.8\columnwidth]{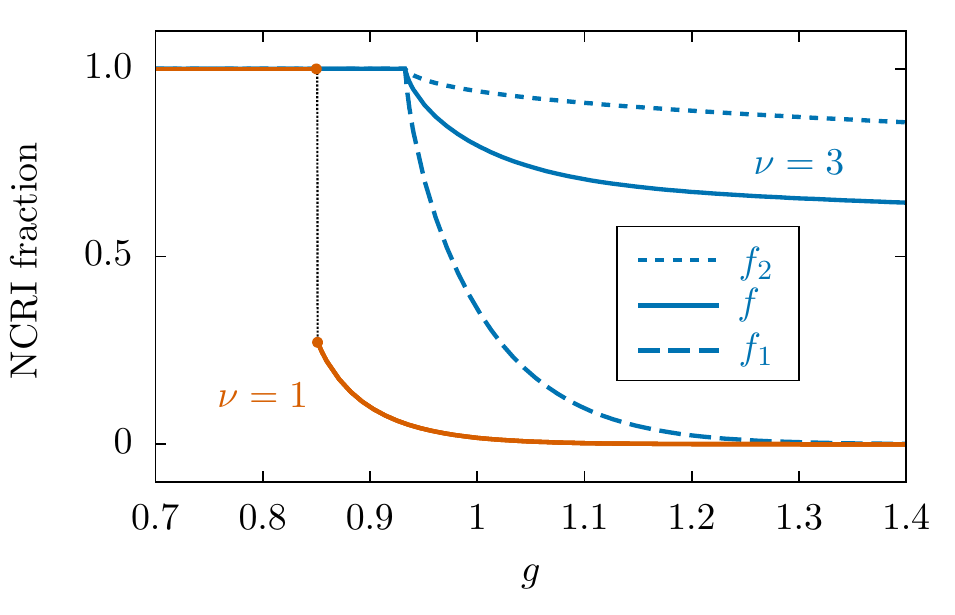}
\caption{({\it Color online}) Non-classical rotational inertia (NCRI) fraction as a function of interaction strength $g$ for $\nu=1$ ({\it orange}; the discontinuity is indicated by the bullets and connecting dashed line), and for $\nu=3$ with $f_1, f_2$ and $f$  as indicated in the legend. (Note that $f=f_1=f_2$ for the symmetric system).}
\label{fig:sff}
\end{figure}

In this Letter, we set focus on the case of asymmetric components confined in a one-dimensional trap with periodic boundary conditions. 
For equal intra-species interactions but  different numbers of particles in the two components, with increasing coupling strength the translational symmetry of the uniform system is broken. A localized droplet forms, stabilized by quantum fluctuations, which coexists with a uniform residual condensate of excess atoms  that cannot bind to the droplet, but are kept together by the confinement. 
We demonstrate that this asymmetric system, although with just a single droplet unlike what is typically seen in 
dipolar supersolids~\cite{Bottcher2021}, simultaneously exhibits rigid-body and superfluid properties. The non-classical rotational inertia (NCRI) reveals that the motion of the droplet at low velocities is not only that of a classical rigid body but is  accompanied by the response of the non-droplet atoms moving in a direction {\it opposite} to the motion of the rigid body. Importantly, this response is found to exist for infinitesimal rotations,
thus having a profound impact on the degree of superfluidity of the system.
The number of atoms contributing to the formation of a vortex is larger than that of the residual condensate, coinciding with the NCRI fraction. Our findings are corroborated by an analysis of  the lowest excitation modes in the compound system.  \\
The energy density for a uniform binary Bose-Bose mixture in one dimension with equal masses and short-ranged interactions, including  BMF corrections, equals~\cite{Petrov2016}
\begin{equation}\label{edens}
\mathcal{E} = \frac{g}{2}(n_1^2 + n_2^2) + (\delta g - g)n_1 n_2 - \frac{2m^{1/2} g^{3/2}}{3\pi \hbar}(n_1+n_2)^{3/2},
\end{equation}
\noindent where $n_i$ are the densities of each component and $m$ the mass of a single particle. Here we have set the intraspecies coupling constants to be equal, $g_{11}=g_{22}=g$, and introduced $\delta g = g + g_{12}$, where $g_{12}$ is the interspecies coupling constant. The first two terms in Eq.~(\ref{edens}) constitute the MF energy density and the last term accounts for the first correction beyond mean field. The energy density Eq.~(\ref{edens}) is valid provided $\eta = \sqrt{mg/n\hbar^2}\ll 1$, ensuring weak interactions (here we have assumed similar order of magnitudes for the densities $n_1\sim n_2 \sim n$), and that $\delta g$ is small in the sense $\delta g/g \sim \eta$. For a finite-size system the extended coupled Gross-Pitaevskii equations corresponding to the energy density Eq.~(\ref{edens}) are
\begin{equation}\label{gpe}
\begin{aligned}
\mathrm{i}\hbar \partial_t \psi_i = &-\frac{\hbar^2}{2m}\partial_{xx}\psi_i + g |\psi_i|^2 \psi_i + (\delta g - g)|\psi_j|^2 \psi_i \\
&- \frac{m^{1/2} g^{3/2}}{\pi \hbar}(|\psi_i|^2 + |\psi_j|^2)^{1/2} \psi_i~,
\end{aligned}
\end{equation}
\noindent where $i,j=1,2$ and $i\neq j$. We impose periodic boundary conditions $\psi_i(x) = \psi_i(x+2\pi R)$, enforcing a  confinement 
of length $2\pi R$. This is a good approximation for a ring of radius $R$ whenever bending effects may be neglected, {\it i.e.} when the transversal confinement length is much smaller than $R$.
(Such binary 1D ring systems have been extensively studied, both experimentally~\cite{Ryu2007,Ramanathan2011,Wright2013,Ryu2013,Beattie2013,Eckel2014,Guo2020,Nicolau2020} and theoretically
~\cite{Smyrnakis2009,Smyrnakis2012,Anoshkin2013,Smyrnakis2014,Abad2014,Mateo2015,Roussou2018,Chen2019,Polo2019,Ogren2021}). \\
From here on, we use dimensionless units such that $\hbar=m=R=1$. The order parameter is normalized to the number of particles in each component according to $\int_0^{2\pi} |\psi_i(x)|^2 \mathrm{d}x = N_i$. The ground state is obtained by solving Eq.~(\ref{gpe}) numerically with a split-step Fourier method in imaginary time. To analyze the system in a rotating frame the term $-\Omega \hat{L}\psi _i$ is added to the right side of Eq.~(\ref{gpe}), where $\hat{L}=-\mathrm{i}\partial_x$, which is then solved in the same manner. In order to find the ground state for a fixed value of the angular momentum $L=L_1+L_2$, where $L_i = \int_0^{2\pi} \psi^\ast_i \hat{L} \psi_i \mathrm{d} x $, we consider the quantity $\Tilde{E} = E + (C/2)(L-L_0)^2$ \cite{Komineas2005}, where $E$ is the total energy corresponding to Eq.~(\ref{gpe}). By minimizing $\Tilde{E}$ for sufficiently large values of the constant $C>0$, the obtained ground state will have angular momentum $L\approx L_0$ since the critical point of $\Tilde{E}$ is $L = L_0 - \partial_L E /C$, which is a minimum whenever $C > -\partial_{LL}E$.
Introducing $\lambda = \delta g/g$, $N = N_1 + N_2$ and $\nu = N_2/N_1$, we illustrate our findings below by fixing $\lambda = 0.01$ and $N = 5000$. The asymmetry parameter is restricted to $\nu\geq 1$. \\
We  first investigate  the density distributions of the two components, starting with the symmetric case where both components of the mixture are equal, $\nu=1$. The upper panel of Fig.~\ref{fig:dens} shows the corresponding densities for different values of the interaction parameter $g$ for a slowly rotating system.
For high enough values of $g$ the ground state has the form of a droplet in agreement with previous findings in one dimension~\cite{Petrov2016, Astrakharchik2018}. As $g$ is decreased the extent of the droplet increases, which results in a transition to a uniform state. We understand this by considering the bulk density of a flat-top droplet $n_0 = 8g/(9\pi^2 \lambda^2)$~\cite{Petrov2016}, which implies a droplet extent $\sim N/(2n_0)$. When this droplet size is much smaller than the ring length the periodic boundary conditions do not significantly affect the system. As $g$ is decreased, the droplet size eventually becomes comparable to the ring length for $g\sim 9\pi N\lambda^2/32$. This results in a transition from a droplet to a uniform system and is a consequence of the periodic boundary conditions. (See also the discussion of boundary effects in~\cite{Cui2021}). For low values of $g$, the situation is similar also in the asymmetric case, where both components display uniform behavior, as shown for $\nu = 3$ in the lower panel of Fig.~\ref{fig:dens}. With increasing interaction strength $g$ the translation symmetry is eventually broken, and the ground state density for the component with more particles changes to what appears to be a mix between a droplet and a uniform medium, while the other component displays a normal droplet solution. The  droplet coexists with a uniform background, since the excess particles in the second component can not bind to it~\cite{Petrov2015}. Deviating from $\nu=1$ results in an increase in the ratio of MF to BMF energy, eventually causing the former to dominate the total energy. Thus, unlike the symmetric case where droplet formation takes place where the MF and BMF terms are of similar orders of magnitude, for the asymmetric system we have BMF effects such as displayed in Fig.~\ref{fig:dens} even though the MF contribution can be much larger than the BMF one. We note that solving Eq.~(\ref{gpe}) without the 
BMF contribution ceteris paribus leads to uniform solutions in the regimes where non-uniform ones were obtained with the full Eq.~(\ref{gpe}) for our choice of parameters.\\
We are here interested in the properties of this mixed-phase system. The ground-state energy $E(\ell )$ as a function of the angular momentum per 
particle $\ell $ is shown 
in Fig.~\ref{fig:disp} for different degrees of asymmetry $\nu$. It takes the form of a single parabola for $\nu=1$, corresponding to the rigid-body rotation of the droplet in the ring.  
For the asymmetric case with $\nu>1$,  different  parabolae appear to intersect. Intriguingly, the structure of $E(\ell )$ is similar to the one found for a dipolar toroidal system in three dimensions in the supersolid phase~\cite{NilssonTengstrand2021}. With this analogy in mind, we  model the system by considering $N_c$ particles taking angular momentum as a classical rigid body under rotation and $N_v$ particles taking angular momentum only in terms of vorticity, with $N = N_c + N_v$. 
The energy cost for adding a  vortex with $s$-fold quantization to the condensate can be determined by assuming an order parameter for the vortex component on the form $\propto \hbox{e}^{\mathrm{i}s x}$, where the integer $s$ is the angular momentum per particle in the  vortex-carrying component, leading to an energy cost equal to $N_v s^2/2$. Since the solid-like part of the system carries angular momentum according to $N_c\ell_c^2/2$, where $\ell_c$ is its angular momentum per particle and $N\ell = N_c\ell_c + N_v s$, the different energetic branches dependent on angular momentum can be written
\begin{equation}\label{ebranches}
E_{s}/N = \frac{1}{2}\left[\frac{(\ell-f_v s)^2}{1-f_v}  + f_v s^2\right],
\end{equation}
\noindent where we have defined the fraction of particles carrying vorticity $f_v=N_v/N$. The branches $E_s$ and $E_{s+1}$ intersect at $\ell = s + 1/2$, {\it i.e.}, having no vortex is energetically favorable for $\ell~<~0.5$, having a singly-quantized one for $0.5 < \ell < 1.5$ and so on. The energies in Eq.~(\ref{ebranches}) obtained within our model predict that the functional form of the dispersion relation should be parabolae intersecting at half-integer values of $\ell$, in accordance with the numerical results displayed in Fig.~\ref{fig:disp}. 
Importantly, this suggests that the asymmetric system can exhibit properties of a 
solid and superfluid simultaneously, with rigid-body rotation and quantized vorticity coexisting. Unlike ring-shaped dipolar supersolids, which display a similar behavior under rotation~\cite{Roccuzzo2019,NilssonTengstrand2021}, here in the case of isotropic short-range interactions 
the ground state density does not exhibit any repeating crystalline structure.
Since the branches $E_s$ have minima at $\ell = f_v s$, there is a 
possibility for the system to exhibit metastable superflow. These minima exist in the ground state energy $E(\ell )$  only if they occur on the interval where the corresponding branch $E_s(\ell )$ is the lowest in energy, thus giving a criterion for the existence of a metastable persistent current related to an $s$ times multiply-quantized vortex according to $f_v>(2s-1)/(2s)$.\\
It is  tempting to identify the amount of particles belonging to the vortex component with the number $N_2-N_1$, intuitively imagining that $N_1$ particles from each component bind as a droplet and thus act in a solid-body fashion while the excess particles take the role of a background superfluid capable of carrying vorticity. If this were true, then $f_v=(N_2-N_1)/N=(\nu-1)/(\nu+1)$, in contradiction with the numerically obtained positions of the minima in Fig.~\ref{fig:disp} which were predicted to occur at $\ell=f_v s$.  To give an example, for $\nu=3$ the first minimum occurs at $\ell\approx 0.63$ while $(\nu-1)/(\nu+1)=0.5$. To investigate this obvious discrepancy, we compute the non-classical rotational inertia fraction for each component~\cite{Leggett1998}
\begin{equation}\label{sffdef}
f_i = 1 - \lim_{\Omega\to 0}\frac{L_i}{N_i\Omega}~.
\end{equation}
\noindent The total NCRI fraction is defined as $f = (N_1f_1 + N_2f_2)/N$, and we plot the numerically obtained results in Fig.~\ref{fig:sff} for $\nu=1$ and $\nu=3$. Interestingly, for $\nu=1$ there is a discontinuous jump from $f=1$ to a value that is non-zero, signaling that even for the symmetric droplet system there are parameter values such that the motion is not entirely classical. As $\nu$ is increased the discontinuity decreases until it eventually disappears, as exemplified for $\nu=3$ in Fig.~\ref{fig:sff}. We see that the total NCRI fraction for the asymmetric system not only differs from $(\nu-1)/(\nu+1)$ but is also dependent on $g$. 
Let us now  study the motion of the condensate under rotation by examining the condensate velocity $\phi_i'(x)$ (where $\phi_i$ is the phase of the order parameter) which can be written 
\begin{equation}\label{velocity}
\phi_i'(x) = \Omega + \frac{L_i - N_i\Omega}{2\pi n_i(x)}~,
\end{equation}
see Appendix A. The velocities for some parameters are plotted in Fig.~\ref{fig:dens} where in particular panels (b) and (d) show how the velocity for both components in the symmetric system and the first component in the asymmetric one are equal to $\Omega $ throughout most of the ring, reflecting the solid-like movement.  Away from the bulk of the droplet where the density is small, the second term in Eq.~(\ref{velocity}) becomes significant and the velocity thus deviates from $\Omega$ in a manner depending on $L_i-N_i\Omega$. For small $\Omega$ this difference is negative, resulting in a change of sign for the velocity as displayed in Fig.~\ref{fig:dens}(b) and (d). If the density is negligible in the region where the velocity deviates from $\Omega$ this will have little effect, but if the droplet instead occupies most of the ring, as is the case close to the transition point between the uniform and droplet phases, this results in parts of the system moving in a direction {\it opposite} to the rest of the condensate. The NCRI fraction will consequently differ from zero, explaining why $f$ for the symmetric system and $f_1$ for the asymmetric system do not immediately fall to zero at the transition point. For the second component in the non-uniform asymmetric case the velocity has opposite signs inside and outside the droplet region, indicating that the movement of the droplet is accompanied by a response of the background medium, which moves in the other direction. Since this response exists also for infinitesimal rotations, it affects the results based on the definition in Eq.~(\ref{sffdef}), increasing it compared to if there had been no response flow of the background. 
Curiously, this implies that the interpretation of $f$ as the fraction of particles that stay at rest as the container is set to slowly rotate is not a correct one. 
(Interestingly, a similar type of response by the background has been noticed also for dipolar condensates in the supersolid phase~\cite{Roccuzzo2020,NilssonTengstrand2021}). To connect to the results in Fig.~\ref{fig:disp}, we compute the corresponding NCRI fractions and find that $f\approx 0.47$, $f\approx 0.63$, $f\approx 0.72$ and $f\approx 0.78$ for $\nu = 2,3,4,5$, respectively. These data agree well with  the positions of the minima of $E(\ell )$, suggesting that $f_v \approx f$, {\it i.e.}, that the fraction of particles related to vortex formation is larger than that of the residual condensate and coincides with the NCRI fraction.
Finally, we investigate the spectra of collective excitations. Following the usual procedure we linearize the extended
Gross-Pitaevskii equation Eq.~(\ref{gpe}) around the ground state $\psi_{0,i}$ and write
\begin{equation}
\psi_i(x,t) = \mathrm{e}^{-\mathrm{i}\mu_i t}\left[\psi_{0,i}(x) + u_i(x)\mathrm{e}^{-\mathrm{i}\omega t} + v_i^\ast(x)\mathrm{e}^{\mathrm{i}\omega t} \right]~,
\end{equation}
\noindent  keeping terms up to first order in the Bogoliubov amplitudes $u_i$ and $v_i$. Here $\mu_i$ is the chemical potential and $\omega/(2\pi)$ is the frequency of oscillation of $u_i$ and $v_i$. Due to the periodic boundary conditions imposed on the system, we expand the amplitudes in plane waves according to $u_i(x) = \mathrm{e}^{\mathrm{i}kx}u_{i,k}(x)$ and $v_i(x) = \mathrm{e}^{\mathrm{i}kx}v_{i,k}(x)$ allowing us to solve the resulting Bogoliubov-de Gennes equations for fixed $k$ (see Appendix B for more details). Fig.~\ref{fig:bdg} shows the lowest modes as a function of $k$ for the symmetric ($\nu=1$) and asymmetric case (here for $\nu=4$) in the non-uniform regime. 
In both cases there are three gapless modes, which we characterize by looking at the density and phase fluctuations of the system, $\Delta n_{i,k} = |u_{i,k}+v_{i,k}|^2$ and $\Delta \phi_{i,k} = |u_{i,k}-v_{i,k}|^2$, respectively \cite{Wu1996,Roccuzzo2019}. Two of these modes are imaginary, indicating a dynamical instability of the system. As $g$ is decreased and we approach the uniform regime these modes harden, implying a higher degree of instability closer to the phase boundary. With increased $g$ the modes instead soften until they become vanishingly small and the system thus becomes stable. The softer of the imaginary modes is at $k=0$ a phase mode, and remains so for $\nu=1$ at all $k$. For $\nu=4$, the characteristics of this mode changes as $k$ is increased, and the component with more particles is instead dominated by fluctuations in the density while the component with fewer particles still is related to fluctuations in the phase. The harder of the imaginary modes is in all cases a density mode, corresponding to a center of mass excitation of the droplet. 
\begin{figure}[H]
\centering
\includegraphics[width = \columnwidth]{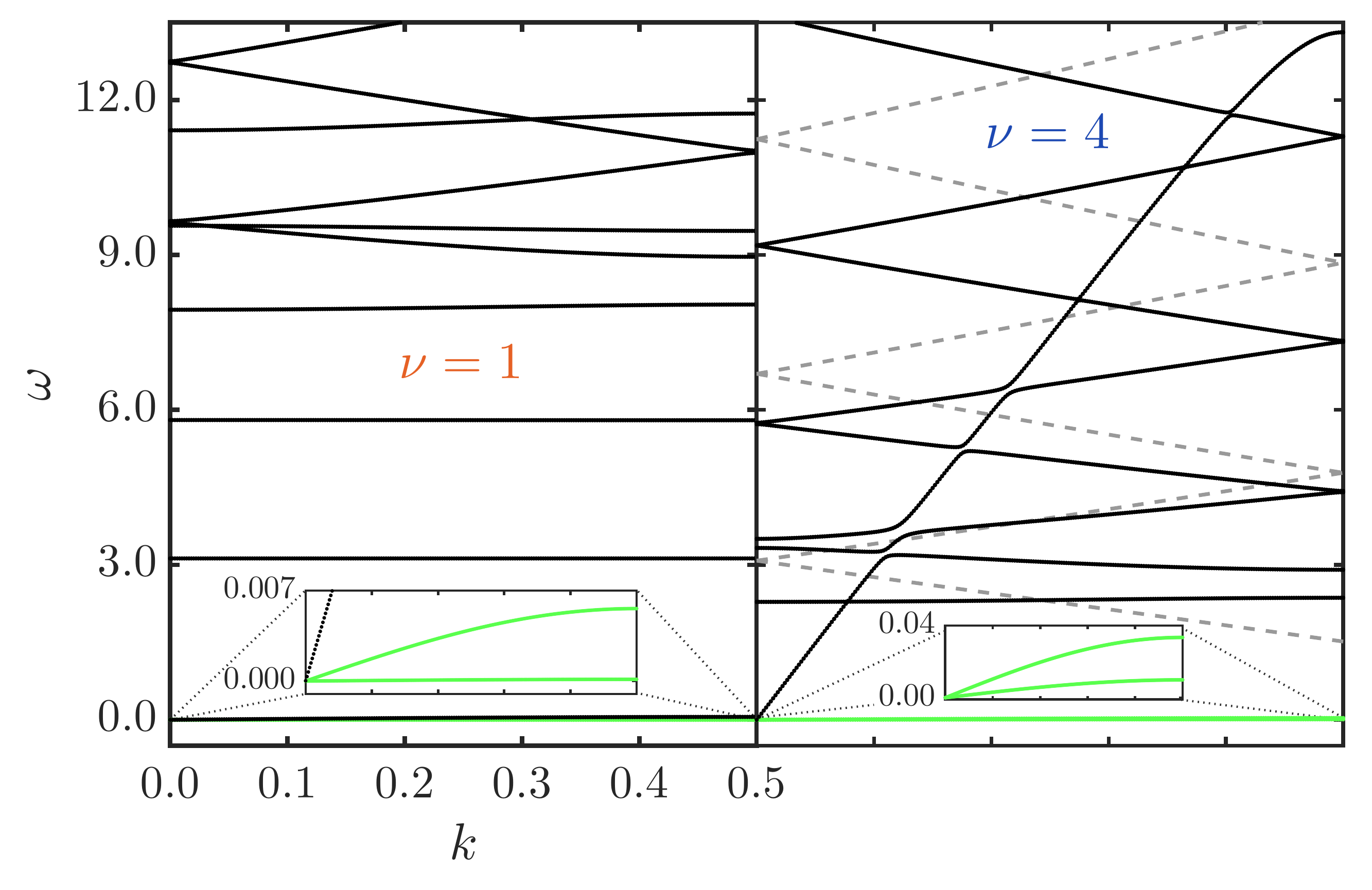}
\caption{{\it (Color online)} Spectra of elementary excitations  for $\nu=1$ (left panel) and $\nu=4$ (right panel) as a function of $k$ for $g=1.4$. The inset shows a zoom-in of the two lowest {\it (green)} modes (which are purely imaginary). The (gray) dashed lines show the analytical solutions in the homogeneous regime for $|p|\geq3$ as discussed in the text for the corresponding parameter values.}
\label{fig:bdg}
\end{figure}
The single real gapless mode is a phase mode that is related to the superfluidity of the system. For the symmetric case, it becomes soft rapidly as the NCRI  goes to zero.  With increasing $\nu$ the superfluid character of the condensate is enhanced, hardening  this mode such that it may no longer be the lowest-lying one, resulting in both crossings and avoided crossings with the gapped modes, see the right panel of Fig.~\ref{fig:bdg}. The lowest-lying gapped modes of the pure droplet system are characterized by their insensitivity to $k$, the first one being the breathing mode~\cite{Tylutki2020}. For the asymmetric system we find that the lowest gapped mode is still the breathing mode, followed by higher modes with linear slopes that form a zig-zag pattern higher up in the spectrum. We can understand this structure from the solution of the Bogoliubov-de Gennes equations of the uniform condensate, which may be obtained analytically (see Appendix B). In this case the low-lying part of the spectrum is dominated by solutions on the form $\omega_{pk} \approx p^2/2 + G + pk$, where $p$ is an integer and $G$ a number that depends only on the interaction strengths and particle numbers but not on $p$ or $k$. Figure~\ref{fig:bdg} clearly shows how the non-uniform system  exhibits a structure that is similar to the uniform one, but displaced to higher energies. This is a clear indication for the increased importance of the non-droplet superfluid background  higher up in the excitation spectrum.\\
In conclusion, we have studied the effects of quantum fluctuations beyond just droplet formation of a one-dimensional binary Bose mixture with short-range interactions in a 1D confinement with periodic boundary conditions. Although the BMF contribution for asymmetric configurations may be small compared to the MF one, it can cause a transition to a system where translational symmetry is broken. Examining the rotational properties and collective excitations  
we found that droplet and superfluid characteristics co-exist. 
Through energetic considerations and from calculating the non-classical rotational inertia we demonstrated 
that the non-droplet part may not be simply identified as the total superfluid, due to its reaction to the droplet's movement. 
Investigating the collective excitations revealed the existence of a single real gapless phase mode that softens with decreased NCRI, tying it to the degree of superfluidity of the system. Further superfluid signatures were observed in the spectrum of the non-uniform asymmetric system at higher energies where the structure became qualitatively similar to that of a homogeneous one. Although the present study focused on 1D, the results are expected to hold also in the quasi-1D limit for a sufficiently strong transversal trapping.
As mentioned above, droplets have been experimentally observed for binary $^{39}$K \cite{Cabrera2018, Semeghini2018}, or hetero-nuclear mixtures~\cite{Errico2019,ZGuo2021}, and have also been confined in a quasi-1D optical wave\-guide~\cite{Cheiney2018}. A ring-shaped superfluid similar to what was envisioned here should thus be straight-forward to realize experimentally. 
Annular flow with long lifetime was realized in anharmonic traps~\cite{YGuo2020}, and ring traps with widely tunable parameters were recently reported in Ref.~\cite{deGoerdeHerve2021}, although for a single-component BEC. In the light of these recent experiments, the realization of 
toroidal droplet-superfluid compounds is a realistic endeavor, opening up for new insight regarding the coexistence of solid-like and superfluid properties. 

{\it Acknowledgments.}
This research was financially supported by the Knut and Alice Wallenberg Foundation, and the Swedish Research Council. 


%

\newpage

\appendix

\section{Condensate Velocity}

The time-independent coupled extended Gross-Pitaevskii equations in the rotating frame are

\begin{equation}\label{tigp}
\begin{aligned}
\mu_i\psi_i = &-\frac{1}{2}\psi''_i + g |\psi_i|^2 \psi_i + (\delta g - g)|\psi_j|^2 \psi_i \\
&- \frac{g^{3/2}}{\pi}(|\psi_i|^2 + |\psi_j|^2)^{1/2} \psi_i - \Omega \hat{L}\psi_i,
\end{aligned}
\end{equation}
\noindent where $\mu_i$ is the chemical potential and it is assumed that $i\neq j$ as in the main text. Writing $\psi_i(x) = \tilde{\psi}_i(x)\mathrm{e}^{\mathrm{i}\phi_i(x)}$ and inserting into Eq.~(\ref{tigp}) we find

\begin{equation}\label{tigp2}
\begin{aligned}
\mu_i\tilde{\psi}_i = &-\frac{1}{2}\tilde{\psi}_i'' + \frac{1}{2}\tilde{\psi}_i(\phi_i')^2+ g \tilde{\psi}_i^3 + (\delta g - g)\tilde{\psi}_j^2 \tilde{\psi}_i \\
&- \frac{g^{3/2}}{\pi}(\tilde{\psi}_i^2 + \tilde{\psi}_j^2)^{1/2} \psi_i - \Omega\tilde{\psi}_i\phi_i' \\
&+\mathrm{i}\left(-\tilde{\psi}_i'\phi_i' - \frac{1}{2}\tilde{\psi}_i\phi_i'' +\Omega\tilde{\psi}_i'\right).
\end{aligned}
\end{equation}

\noindent Using the fact that $\tilde{\psi}_i$ and $\phi_i$ are real we separate Eq.~(\ref{tigp2}) into two equations by identifying real and imaginary terms, where the imaginary part yields the equation

\begin{equation}
\frac{1}{2}\tilde{\psi}_i\phi_i'' + \tilde{\psi}_i'\phi_i' = \Omega\tilde{\psi}_i'.
\end{equation}

\noindent This equation can be solved for $\phi_i'(x)$ in terms of $\tilde{\psi}_i(x)$, with the solution

\begin{equation}
\phi_i'(x) = \Omega + \frac{C_i}{n_i(x)},
\end{equation}

\noindent where $C_i$ is an integration constant and $n_i=~|\psi_i|^2=~\tilde{\psi}_i^2$. Since the angular momentum can be written $L_i = \int  n_i(x) \phi_i'(x) \mathrm{d}x$ we finally obtain

\begin{equation}
\phi_i'(x) = \Omega + \frac{L_i - N_i\Omega}{2\pi n_i(x)},
\end{equation}

\noindent which is the expression for the condensate velocity used in the main text.

\section{Bogoliubov-de Gennes Equations}

The extended Gross-Pitaevskii equation Eq.~(4) of the main text is linearized around the ground state $\psi_{0,i}$ by writing 
\begin{equation}
\psi_i(x,t) = \mathrm{e}^{-\mathrm{i}\mu_i t}\left[\psi_{0,i}(x) + u_i(x)\mathrm{e}^{-\mathrm{i}\omega t} + v_i^\ast(x)\mathrm{e}^{\mathrm{i}\omega t} \right]
\end{equation}
\noindent and keeping terms up to first order in the Bogoliubov amplitudes $u_i$ and $v_i$ (here $\omega/2\pi$ is the frequency of oscillation of $u$ and $v$). Due to the periodicity of the system we expand the amplitudes in plane waves $u_i(x) = \mathrm{e}^{\mathrm{i}kx}u_{i,k}(x)$ and $v_i(x) = \mathrm{e}^{\mathrm{i}kx}v_{i,k}(x)$, resulting in the Bogoliubov-de Gennes equations
\begin{equation}\label{bdgeqs}
\mathbf{A}_k\mathbf{v}_k = \omega_k \mathbf{v}_k,
\end{equation}
\noindent where $\mathbf{v}_k = \begin{pmatrix}u_{1,k}(x) & v_{1,k}(x) & u_{2,k}(x) & v_{2,k}(x)\end{pmatrix}^\mathrm{T}$ and
\begin{equation}
\mathbf{A}_k = 
\begin{pmatrix}
X_{12,k} & Y_{1} & Z & Z \\
-Y_{1} & -X_{12,k} & -Z & -Z \\
Z & Z & X_{21,k} & Y_{2} \\
-Z & -Z & -Y_{2} & -X_{21,k}
\end{pmatrix}
\end{equation}
\noindent with
\begin{equation}
\begin{aligned}
X_{ij,k} &= -\frac{1}{2}\frac{\partial^2}{\partial x^2} -\mathrm{i}k\frac{\partial}{\partial x} + \frac{k^2}{2} - \mu_i \\
&+ (2g-3\alpha)\psi_{0,i}^2 + (\delta g-g-2\alpha)\psi_{0,j}^2 \\
Y_{i} &= (g-\alpha)\psi_{0,i}^2  \\
Z &= (\delta g - g - \alpha)\psi_{0,1}\psi_{0,2}
\end{aligned}
\end{equation}
\noindent and
\begin{equation}
\alpha = \frac{g^{3/2}}{2\pi(\psi_{0,1}^2 + \psi_{0,2}^2)^{1/2}}.
\end{equation}
\noindent Here $\alpha$ is the contribution due to beyond mean-field effects and we have assumed $\psi_{0,i}$ to be real. In the homogeneous regime the BdG equations can be solved analytically by writing \cite{Gao2019}
\begin{equation}\label{ansatz}
\mathbf{v}_k(x)
= 
\sum_{p=-\infty}^\infty \frac{e^{\mathrm{i} p x}}{\sqrt{2\pi}}
\tilde{\mathbf{v}}_{pk},
\end{equation}
\noindent where $\mathbf{\tilde{v}}_{pk} = \begin{pmatrix} \tilde{u}_{1,pk} & \tilde{v}_{1,pk} & \tilde{u}_{2,pk} & \tilde{v}_{2,pk}\end{pmatrix}^\mathrm{T}$. By substituting into Eq.~(\ref{bdgeqs}) it can be seen that $\mathrm{e}^{\mathrm{i}p x}\tilde{\mathbf{v}}_{pk}/\sqrt{2\pi}$ is an eigenvector with eigenvalue $\omega_{pk}$ satisfying
\begin{equation}\label{bdgeqshom}
\tilde{\mathbf{A}}_{pk}\tilde{\mathbf{v}}_{pk} = \omega_{pk} \tilde{\mathbf{v}}_{pk},
\end{equation}
\noindent where
\begin{equation}
\tilde{\mathbf{A}}_{pk} = 
\begin{pmatrix}
\tilde{X}_{1,pk} & Y_{1} & Z & Z \\
-Y_{1} & -\tilde{X}_{1,pk} & -Z & -Z \\
Z & Z & \tilde{X}_{2,pk} & Y_{2} \\
-Z & -Z & -Y_{2} & -\tilde{X}_{2,pk}
\end{pmatrix},
\end{equation}
\noindent and $\tilde{X}_{i,pk} = (p+k)^2/2 + (g-\alpha)n_i$ with $n_i = N_i/(2\pi)$. The solutions to Eq.~(\ref{bdgeqshom}) are
\begin{equation}
\begin{aligned}
&\omega_{\pm,pk}^2= \frac{(p+k)^2}{4\pi} \bigg[\pi(p+k)^2 + (g-\alpha)N \\
&\pm \sqrt{(g-\alpha)^2(N_1-N_2)^2 + 4(\delta g - g - \alpha)^2 N_1 N_2}\bigg],
\end{aligned}
\end{equation}
\noindent which in the limit $g \gg \delta g, \alpha$ can be written
\begin{equation}\label{omp}
\omega_{+,pk}^2 \approx \frac{(p+k)^2}{4\pi} \bigg[\pi (p+k)^2 + 2(g-\alpha)N - 4(\delta g - 2\alpha)\frac{N_1 N_2}{N} \bigg]
\end{equation}
\noindent and
\begin{equation}\label{omm}
\omega_{-,pk}^2 \approx \frac{(p+k)^2}{4\pi} \bigg[\pi (p+k)^2 + 4(\delta g - 2\alpha)\frac{N_1 N_2}{N} \bigg].
\end{equation}
From these two types of solutions we observe that there is a separation of scales, where the low-lying part of the spectrum is dominated by the $\omega_{-,pk}$ branches. For sufficiently large $|p|$, $k\in[0.0,0.5]$ and $\omega_{-,pk} > 0$, we have the approximate form
\begin{equation*}
\omega_{-,pk} \approx \frac{p^2}{2} + (\delta g - 2\alpha)\frac{N_1 N_2}{\pi N} + pk,
\end{equation*}
\noindent which on the interval $k\in[0.0,0.5]$ takes the form of a zig-zag pattern as discussed in the main text and displayed in Fig.~5 of the paper.

\end{document}